\documentclass[runningheads]{llncs}
\usepackage[T1]{fontenc}
\usepackage[utf8]{inputenc}
\usepackage{lmodern}
\usepackage{amsmath, amssymb}
\usepackage{enumitem}
\usepackage{geometry}
\usepackage{hyperref}
\usepackage{graphicx}
\begin{document}
\title{Towards an ABM on Proactive Community Adaptation\\for Climate Change\thanks{The work reported here is part of the EU Horizon PRO-CLIMATE project with grant agreement No. 101137967.}}
%
%\titlerunning{Abbreviated paper title}
% If the paper title is too long for the running head, you can set
% an abbreviated paper title here
%
\author{
\"{O}nder G\"{u}rcan\inst{1}\orcidID{0000-0001-6982-5658} \and
David Eric John Herbert\inst{2}\orcidID{0000-0002-9551-6514} \and
F. LeRon Shults\inst{1,3}\orcidID{0000-0002-0588-6977} \and
Christopher Frantz\inst{4}\orcidID{0000-0002-6105-8738} \and
Ivan Puga-Gonzalez\inst{1}\orcidID{0000-0003-2510-6760}
}

%\author{\IEEEauthorblockN{Anonymous Authors}}

%
\authorrunning{\"{O}. G\"{u}rcan et al.}
% First names are abbreviated in the running head.
% If there are more than two authors, 'et al.' is used.
%

\institute{
    Center for Modeling Social Systems,
    NORCE Norwegian Research Center AS, Norway\\
    \and
    Department of Sociology, University of Bergen, Norway\\
   \and
    Institute for Global Development and Planning, University of Agder, Norway\\
    \and
    Department of Computer Science, Norwegian University of Science and Technology (NTNU), Norway
    \\
    \email{ongu@norceresearch.no}
}
\maketitle              % typeset the header of the contribution
\begin{abstract}
We present an agent-based model (ABM) simulating proactive community adaptation to climate change in an urban context. The model is applied to Bergen, Norway, represented as a complex socio-ecological system. It integrates multiple agent types: municipal government (urban planners and political actors), civil society (individual citizens), environmental NGOs and activists, and media. Agents interact during urban planning processes—particularly the evaluation and approval of new development proposals. Urban planners provide technical assessments, while politicians (organized by party) make final decisions to approve, modify, or reject projects. Environmental NGOs, activist groups, and the media shape public perception and influence policymakers through campaigns, lobbying, protests, and news coverage. Individual citizens decide whether to engage in collective action based on personal values and social influences. The model captures the resulting decision-making ecosystem and reveals feedback loops and leverage points that determine climate-adaptive outcomes. By analyzing these dynamics, we identify critical intervention points where targeted policy measures can facilitate systemic transformation toward more climate-resilient urban development.

\end{abstract}
\section{Introduction}

Climate change is a defining global challenge that necessitates proactive adaptation strategies at the community level. The PRO-CLIMATE project addresses this critical issue by fostering social transformation and behavioural change within diverse communities across Europe. Recognizing communities as complex adaptive systems influenced by interconnected socio-economic and environmental factors, PRO-CLIMATE aims to identify social tipping points and actionable policy interventions to facilitate systemic transformation towards climate resilience.

Central to the project’s methodology is an innovative agent-based modelling (ABM) approach, designed to realistically represent and analyze European socio-ecological systems. The initial agent-based model described in this abstract represents our foundational step towards understanding the dynamics of behavioural adaptations to climate threats at the community scale. Through interactive simulations, this model aims to capture the complexities of stakeholder interactions, governance structures, and community responses to policy actions.

The model presented here sets the groundwork for further development within PRO-CLIMATE, contributing to the identification of systemic interdependencies, leverage actions, and ultimately informing policy recommendations to accelerate proactive climate adaptation and systemic societal change.

\section{State of the Art}

Agent-based modeling has emerged as a powerful tool for exploring complex socio-ecological systems and has been extensively applied to climate change research, particularly for evaluating the effectiveness of adaptation strategies \cite{Lawyer2023,Bury2019,Beckage2018,Hoekstra2017,Greeven2016,Oviedo2016,Arneth2014,Balbi2010}. 
Current ABM studies typically focus on modeling individual decision-making processes and interactions to understand collective behavioral dynamics under different policy scenarios. 
Such models have successfully illustrated the emergence of cooperation, social norms, and tipping points critical for societal adaptation to environmental stressors.

However, many existing models lack sufficient granularity to represent the diverse socio-economic contexts across Europe comprehensively.
Moreover, limited integration of real-world governance structures and stakeholder inputs restricts their applicability and accuracy. 
PRO-CLIMATE addresses these gaps by integrating extensive stakeholder collaboration through living labs, enabling a more realistic representation of community-specific adaptation pathways. 
By enhancing the fidelity of agent interactions and incorporating extensive empirical validation, the proposed model aims to significantly advance the predictive power and practical relevance of ABM approaches in climate adaptation research.

\section{The Agent Modelling Approach}

\begin{figure}[ht!]
    \centering
    %\begin{subfigure}[b]{0.45\textwidth}
         \centering
        \includegraphics[width=0.40\textwidth]{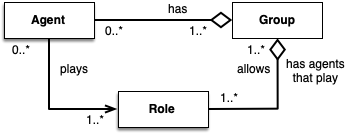}
    %     \caption{organisational Model}
    %     \label{fig:AGR-organisational-Model}
    %\end{subfigure}
    %\hfill
    %\begin{subfigure}[b]{0.45\textwidth}
    %     \centering
    %     \includegraphics[width=0.95\textwidth]{figures/AGR_Cheeseboard.png}
    %     \caption{Cheeseboard organisational diagram}
    %     \label{fig:AGR-Cheeseboard}
    %\end{subfigure}
    \caption{Agent/Group/Role organisational model.}
    %(a) and as a cheeseboard diagram (b).}
    \label{fig:AGR-Representations}
\end{figure}

Among several multi-agent-oriented approaches \cite{Abbas2015,Criado2013,Giorgini2006,Dignum2005,Ferber2004,Hubner2002} proposed in the literature, the Agent/Group/Role (AGR) approach proposed in \cite{Ferber2004} is a good fit for our motivations and purpose. 
In particular, AGR describes what constitutes a Multi-Agent System (MAS) organization at a high level of abstraction and is therefore highly flexible and adaptable to various interaction schemes and organizational designs \cite{Roussille2022}. 
The AGR model (Figure \ref{fig:AGR-Representations}) is based on three first-class abstractions: agent, group, and role.

Roles are abstract representations of functional positions of agents within a group. A role outlines the responsibilities it entails, the constraints agents must satisfy to assume that role, and the benefits agents may gain by performing it.

Groups identify contexts for patterns of activity (i.e., roles) that are shared by groups of agents (i.e., they bring together agents collaborating within a common setting). Agents may communicate if, and only if, they exist within the same group. Groups serve as organizational structures [60] in which interactions form a functionally coherent aggregate of agents. Moreover, groups may also establish boundaries. Agents outside a given group may not be aware of its internal structure.

Agents are active, communicating entities that perform roles within groups. An agent plays at least one role in a group but may take on multiple roles and belong to multiple groups as well. However, the AGR approach imposes no constraints on the internal architectures, cognitive abilities, or mental states of agents.

In our work, we prefer to refer to groups as environments, as this term better conveys the idea of a broader social context or setting in which agents interact, organize, and collaborate.

\section{The Agent-based Model}

\begin{figure}[ht]
    \centering
    \includegraphics[width=0.99\textwidth]{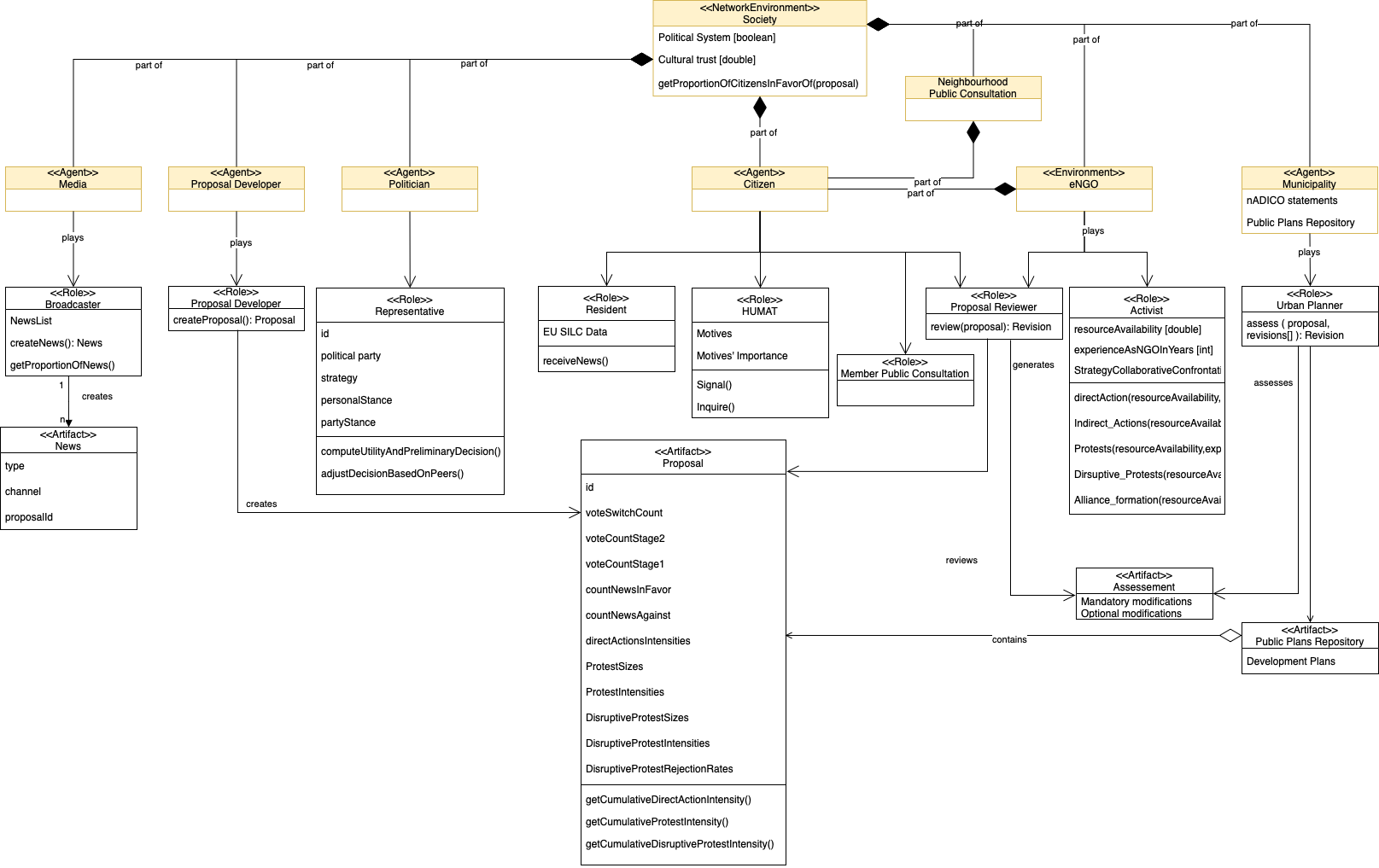}
    \caption{The organizational agent-based model on proactive community adaptation for climate change.}
    \label{fig:agent-based-model}
\end{figure}

The PRO-CLIMATE agent-based model represents the city of Bergen as a complex socio-ecological system composed of interacting agents, roles, environments, and artifacts (see Figure \ref{fig:agent-based-model}). 
The model focuses on the decision-making process surrounding urban development proposals, incorporating governmental structures, civil society dynamics, environmental activism, and media influence. 
Each agent interacts within its corresponding environment and adopts roles that guide its behavior and interactions with others. Below, we describe the key agent types and their roles in the system.

\subsection{Media}

Media agents operate in the Society network environment, assuming the role of Broadcaster. They create news characterized by attributes such as type, impact, reach, and age. Media agents shape public discourse by generating and disseminating news, either independently or in collaboration with activist groups. They influence the public's stance on proposals and indirectly affect policymakers through the modulation of public sentiment.

\subsection{Proposal Developers}

Proposal developers are responsible for initiating new urban development plans. They act through the Proposal Developer role, which generates Proposal artifacts. These proposals become the central object of deliberation across multiple components of the system, progressing through stages of assessment, revision, and decision-making.

\subsection{Politicians}

Political agents operate in the Society environment, taking on the role of Representative. 
They are influenced by their party’s position, cultural trust levels, and the proportion of citizens in favor of specific proposals. 
Politicians make final decisions on development plans and are central to policy formation. Their choices are shaped by assessments from urban planners, citizen sentiment, media coverage, and lobbying efforts by eNGOs and activists.

\subsection{Citizens}

Citizen agents also reside in the Society environment and can adopt multiple roles. As Residents, they bring socio-economic characteristics into the model using EU-SILC data. 
As Members of eNGOs, they participate in organized environmental advocacy. 
Their behavior is driven by the HUMAT model \cite{Jager2025,Gurcan2023}, which evaluates motives (e.g., personal values, social influence, experiences) and informs decisions to signal concern or inquire about proposals. 
Citizens may also take part in public consultations as Proposal Reviewers, producing revisions that feed back into the development process.

\subsection{Activists and eNGOs}

Operating within the eNGO environment, activist agents adopt the Activist role, defined by attributes such as resource availability, experience, and strategic orientation (collaborative or confrontational). They employ various methods: Direct Actions (e.g., lobbying, formal participation), Indirect Actions (e.g., awareness campaigns), Protests (standard and disruptive), and  Alliance Formation.

Their activities aim to influence both public sentiment and governmental decisions. Activists may also participate in proposal review processes, acting as Members of Public Consultation.

\subsection{Municipality and Urban Planners}

Municipal agents operate in the Municipality environment and assume the Urban Planner role. 
They assess proposals and revisions, producing Assessment artifacts that include mandatory and optional modifications. 
These assessments are stored in a Public Plans Repository and play a critical role in informing political decisions. 
Urban planners rely on formal rules, nADICO statements, and policy guidelines to ensure evaluations are consistent and technically sound.

\subsection{Proposal Lifecycle}

Proposals evolve through iterative assessment and consultation:

\begin{enumerate}
    \item \textbf{Creation}: Initiated by a Proposal Developer.
    \item \textbf{Review}: Evaluated by Citizens and Activists during Public Consultations.
    \item \textbf{Assessment}: Formally assessed by Urban Planners.
    \item \textbf{Deliberation}: Influenced by Media, Civil Society, and eNGO activities.
Decision: Final judgment rendered by Politicians.
\end{enumerate}

Each stage generates new artifacts or triggers agent behaviors, contributing to a feedback-rich system of deliberation and influence.
The overall goal of the model is to simulate the process that leads to a final decision on new development proposals—whether they are rejected outright or sent for revision. This final decision rests with politicians who consider a range of inputs: the technical assessments from urban planners, their party's political stance on climate change, the aggregated voices of civil society (including the actions of eNGOs and activists), media-driven public sentiment, and their own political and personal stances toward climate change versus economic growth.

By integrating these components, the model captures the dynamic interplay between government, civil society, activist groups and media. This comprehensive approach provides a robust platform to identify critical leverage points and effective policy strategies.

\section{Conclusion and Future Work}

The initial agent-based model presented in this abstract provides a foundational framework for exploring community-level responses to climate change through socio-behavioural adaptation. 
Building on this initial design, future work within PRO-CLIMATE will involve extensive experimentation \cite{Gurcan2013} to simulate various climate scenarios and intervention strategies. 
Subsequent analyses will focus on evaluating outcomes, refining the model based on empirical validations through living labs, and identifying effective policy measures to achieve transformative societal change. 
Ultimately, these steps will contribute to the development of robust methodologies and tools for proactive climate adaptation across diverse European contexts.

%\subsubsection{Acknowledgements} Please place your acknowledgments at
%the end of the paper, preceded by an unnumbered run-in heading (i.e.
%3rd-level heading).
Disclaimer: Funded by the European Union (grant number 101137967). Views and opinions expressed are, however, those of the author(s) only and do not necessarily reflect those of the European Union or the European Climate, Infrastructure and Environment Executive Agency (CINEA). Neither the European Union nor CINEA can be held responsible for them.

%
% ---- Bibliography ----
%
% BibTeX users should specify bibliography style 'splncs04'.
% References will then be sorted and formatted in the correct style.
%
\bibliographystyle{splncs04} 
\bibliography{references}

\end{document}